\documentstyle[preprint,revtex,eqsecnum]{aps}
\begin{document}
\draft
\preprint{PSU/TH/120}
\begin{title}
Models for local ohmic quantum dissipation
\end{title}

\author{Michael R. Gallis}
\begin{instit}
Department of Physics, Pennsylvania State University,
Schuylkill
Campus,\\
Schuylkill Haven, Pennsylvania 17972\\
Internet: MRG3@psuvm.psu.edu
\end{instit}
\begin{abstract}
        We propose a family of master equations for local
quantum
dissipation.  The
master equations are constructed in the form of Lindblad
generators,
with the constraints
that the dissipation be strictly linear (i.e. ohmic),
isotropic and
translationally invariant.
The resulting master equations are given in both the
Schr\"odinger
and
Heisenberg forms.
We
obtain fluctuation-dissipation relations, and discuss the
relaxation of
average kinetic energy to
effective thermal equilibrium values.  We compare our results
for
one
dimension to the Dekker
master equation [H. Dekker, Phys. Rep. {\bf 80}, 1 (1981)],
which can
be
interpreted as a
low
length scale approximation of our model, as well as the
Caldeira-
Leggett
master equation
[A. O. Caldeira and A. J. Leggett, Physica (Utrecht) {\bf 121
A}, 587
(1983)].
\end{abstract}
\pacs{PACS number 3.65.-w, 3.65.Bz, 5.40.+j}

\narrowtext

\section{Introduction}
\label{sec:intro}

        Dissipative quantum systems and quantum brownian motion
have long
been of
interest and a wide variety of techniques have been used to
investigate
them.
{}~\cite{Dekker,Grabert}, We will not attempt a comprehensive
review,
but
extensive
reviews
can be
found in Refs. \cite{Dekker} and \cite{Grabert}.
For linear dissipation, one wishes to obtain dynamics for a
quantum
mechanical system
which are in a sense generalizations of the classical
equations of
motion for linear
dissipation:
\begin{eqnarray}
m\dot{\bf x} &=&{\bf p},\nonumber \\
\dot{\bf p} &=&-{\eta  \over m}{\bf p} +{\bf F}.
\label{class}
\end{eqnarray}
The classical Langevin equations (and nonlinear
generalizations) can
be
obtained by
considering a particle interacting with an oscillator
bath.~\cite{zwanzig}  Quantum
mechanical
calculations often begin with the same type of environment.
For
example, Caldeira and
Leggett \cite{Caldeira} use the influence functional
technique of
Feynman and
Vernon~\cite{Feynman}
for a particle
linearly coupled to a thermal bath of harmonic oscillators.
The
resulting (high temperature)
master equation is given by
\FL
\begin{equation}
 {\partial \rho  \over \partial t}={1 \over i\hbar}[H,\rho
]+{\eta
\over
i2m\hbar}[x,\{p,\rho \}]-{\eta {k}_{B}T \over
{\hbar}^{2}}[x,[x,\rho ]].
\end{equation}
An alternative approach by Dekker\cite{Dekker}, uses the
quantization of
complex
classical canonical
coordinates and quantal noise operators, resulting in the
master
equation:
\FL
\begin{eqnarray}
{\partial \rho  \over \partial t} &=& {1 \over i\hbar}[H,\rho
]\nonumber
\\
&&-i{\lambda  \over \hbar}[x,\{p,\rho \}]+{1 \over
{\hbar}^{2}}({D}_{xp}+{D}_{px})[x,[p,\rho
]]\nonumber \\
&&-{{D}_{xx} \over {\hbar}^{2}}[p,[p,\rho ]]-{{D}_{pp} \over
{\hbar}^{2}}[x,[x,\rho
]].
\label{dequ}
\end{eqnarray}
Neither the Caldeira-Leggett nor the Dekker master equation
is
translationally invariant,
and so cannot represent the effect on a quantum particle of a
local
interaction with a
spatially homogenous environment.  Caldeira and Leggett have
pointed out
that it is only
strictly linear coupling (i.e. bilinear in system and
oscillator bath
coordinates) that can
produce strictly linear dissipation.  However, it may be true
that
other
forms of coupling
produce linear dissipation as an approximation which is valid
over a
wide range of
conditions.  Moreover, the dynamics of quantum coherence
could be
very
different in a
nonlinear (local coupling) versus a linear
model~\cite{GF,Gallis}.  It
is the intent of this
paper
to show one method of constructing a master equation which
reflects
(in
translational
invariance) the symmetries of the environment and
environment-
system
coupling which
presumably give rise to the dissipative dynamics.

        The approach to constructing the master equation is
essentially
algebraic.  We begin
with a master equation of the Lindblad form:
\begin{eqnarray}
 {\partial \rho  \over \partial t}&=&L[\rho ] \nonumber \\
 &=&{1 \over i\hbar}[H,\rho ]+{1 \over
2\hbar}\sum\nolimits\limits_{\mu }^{} [{V}_{\mu }^{}\rho
,{V}_{\mu
}^{\dagger
}]+[{V}_{\mu }^{},\rho {V}_{\mu }^{\dagger }]\nonumber \\
&=&{1 \over i\hbar}[H,\rho ]+\Delta L[\rho
],
\end{eqnarray}
which, for bounded $H$, $\{V_\mu\}$, generates a completely
positive
dynamical
semigroup.~\cite{Lindblad}  The
generator for the corresponding Heisenberg picture is given
by:
\FL
\begin{eqnarray}
 {dO \over dt}-{\partial O \over \partial t}&=&{L}^{{}^*}[O]
\nonumber
\\
&=&-{1 \over
i\hbar}[H,O]+{1 \over 2\hbar}\sum\nolimits\limits_{\mu }^{}
{V}_{\mu
}^{\dagger
}[O,{V}_{\mu }^{}]+[{V}_{\mu }^{\dagger },O]{V}_{\mu
}^{}\nonumber
\\
&=& -{1 \over i\hbar}[H,O]+\Delta {L}^{{}^*}[O].
\end{eqnarray}
Following an approach used by Sandulescu and Scutaru to
generate
the
Dekker master
equation,~\cite{Sandulescu}  we will operate under the
assumption
that
this is also an
appropriate form for a
completely positive dynamical semigroup for unbounded
operators.
We
will focus on the
additional part of the master equation (Langevin equation)
associated
with dissipation, $\Delta L$
 $(\Delta
L^{{}^*})$.

        In Sec.\ \ref{sec:Langevin}  we will examine the
classical
Langevin equations for
a particle coupled
to a harmonic oscillator bath, which will help clarify the
connection
between first principle
classical and quantum calculations, as well as the issues
which arise
for quantum
generalizations of classical systems.  In  Sec.\
\ref{sec:construct}, we
will construct
$\Delta L$
by choosing a
particular form for the operators $\{V_\mu \}$, and establish
some
constraints on these
operators
from the requirement of linear dissipation.  We will choose
particular
form for the
operators
which satisfy the constraints, yield a master equation which
is
translationally invariant, and
which correspond to an isotropic environment.  The results
are
presented
in both the
Schr\"odinger and Heisenberg forms.  In Sec.\
\ref{sec:dissipation}
we
will derive and
discuss
fluctuation-dissipation relations for these models, as well
as the time
dependence of
expectations of
operators which are functions of position or momentum only.
In
Sec.\
\ref{sec:compare},
we
compare
our results for one dimension to the Dekker and Caldeira-
Leggett
master
equations.  We
will comment on our results in Sec.\ \ref{sec:discuss}.

\section{Classical Langevin Equations}
\label{sec:Langevin}

        In this section, we review some aspects of Langevin
equations
which arise from
interaction with a harmonic oscillator bath, essentially
following
Zwanzig~\cite{zwanzig}.
The
composite system has a lagrangian given by
\FL
\begin{eqnarray}
 L&=&{1 \over 2}m{\dot{x}}^{2}-U({\bf x}) \nonumber \\
&&+\sum\nolimits\limits_{\mu }^{} {{m}_{\mu }
\over 2}[{\dot{q}}_{\mu }^{2}-{\omega }_{\mu }^{2}({q}_{\mu
}-
{a}_{\mu }({\bf x}){)}^{2}]
\end{eqnarray}
where ${\bf x}$ is the position of the system of interest,
and
${q_\mu}$
are the
coordinates of the
oscillators comprising the bath.  The interaction terms are
of the
form
$q_\mu {m}_{\mu} \omega_\mu^2 a_\mu({\bf x})$, and the
additional terms of the form ${1\over 2} m_\mu \omega_\mu^2
a_\mu^2({\bf
x})$ are
associated with
regularization of the
potential.  The equation of motion for an individual
oscillator is given
by
\begin{equation}
{\ddot{ q}}_{\mu }=-{\omega }_{\mu }^{2}{q}_{\mu }+{\omega
}_{\mu
}^{2}{a}_{\mu }({\bf x}),
\end{equation}
which is simply a driven simple harmonic oscillator.  The
formal
solution to the equation
of motion is
\begin{eqnarray}
{ q}_{\mu }(t)&=&{q}_{\mu }(0)\cos{\omega }_{\mu
}t+{{\dot{q}}_{\mu }(0)
\over
{\omega }_{\mu }}\sin{\omega }_{\mu }t \nonumber \\
&&+\int_{0}^{t}\sin{\omega }_{\mu }(t-
s) {\omega }_{\mu }{a}_{\mu }({\bf x}(s))ds
\end{eqnarray}
This becomes, after an integration by parts,
\begin{eqnarray}
{q}_{\mu }(t)=&&{q}_{\mu }(0)\cos{\omega }_{\mu
}t+{{\dot{q}}_{\mu }(0) \over
{\omega }_{\mu }}\sin{\omega }_{\mu }t+{a}_{\mu }(x(t)
\nonumber
\\
&-&\cos{\omega }_{\mu }(t-s) {a}_{\mu }(x(0)) \nonumber \\
&-&\int_{0}^{t}\cos{\omega }_{\mu }(t-s){\partial {a}_{\mu
}({\bf
x}(s))
\over \partial
{x}_{j}}{\dot{x}}_{j}ds .
\label{qmotion}
\end{eqnarray}
The equation of motion for the particle is given (for the
component
$x_i$) by
\FL
\begin{equation}
{\ddot{ x}}_{i}(t)=-{\partial U \over \partial
{x}_{i}}+\sum\nolimits\limits_{\mu
}^{} {m}_{\mu }{\omega }_{\mu }^{2}({q}_{\mu }-{a}_{\mu
}({\bf
x})){\partial
{a}_{\mu }({\bf x}) \over \partial {x}_{i}}.
\label{xmotion}
\end{equation}
Direct substitution of equation Eq.\ (\ref{qmotion}) into
Eq.\
(\ref{xmotion}) yields
 an equation for
motion for {\bf x} which can be written
\begin{eqnarray}
{\ddot{ x}}_{i}(t)&=&-{\partial U({\bf x}) \over \partial
{x}_{i}}+{f}_{i}({\bf x}
,t) \nonumber \\
&&-\int_{0}^{t}{\eta }_{ij}({\bf x}(t),{\bf x}(s);t-
s){\dot{x}}_{j}(s)
ds,
\end{eqnarray}
(summation over $j$ implied) where ${\bf f}({\bf x},t)$ is
interpreted
as a fluctuating force given
by
\begin{eqnarray}
 {f}_{i}({\bf x},t) = \sum\nolimits\limits_{\mu}^{} &\{&
{m}_{\mu
}{\omega }_{\mu
}^{2}[({q}_{\mu }(0)-{a}_{\mu }({\bf x}(0))\cos{\omega }_{\mu
}t \nonumber \\
&&+{{\dot{q}}_{\mu }(0) \over {\omega }_{\mu }}\sin{\omega
}_{\mu
}t]{\partial
{a}_{\mu }({\bf x}(t)) \over \partial {x}_{i}}\},
\end{eqnarray}
and the nonlinear dissipation kernel is given by
\FL
\begin{eqnarray}
 {\eta }_{ij}({\bf x}(t),{\bf x}(s);t-s)= \nonumber \\
\sum\nolimits\limits_{\mu }^{}
{m}_{\mu }{\omega }_{\mu }^{2}{\partial {a}_{\mu }({\bf
x}(t)) \over
\partial
{x}_{i}}{\partial {a}_{\mu }({\bf x}(s)) \over \partial
{x}_{j}}\cos{\omega }_{\mu
}(t-s).\end{eqnarray}

The initial probability distribution of the oscillators
determines
the statistics of the fluctuating force.  The oscillators are
assumed to
be
 independent and in a thermal state, with
\begin{equation}
 \langle {q}_{\mu }(0)-{a}_{\mu }({\bf x}(0))\rangle =\langle
{\dot{q}}_{\mu
}(0)\rangle =0,
\end{equation}
and
\FL
\begin{eqnarray}
\langle {1 \over 2}{m}_{\mu }{\omega }_{\mu }^{2}({q}_{\mu
}(0)-
{a}_{\mu
}({\bf x}
(0)))^{2}\rangle &=&{1 \over 2{m}_{\mu }}\langle
({\dot{q}}_{\mu
}(0){)}^{2}\rangle
\nonumber \\
 &=&{1 \over 2}{k}_{B}T.
\end{eqnarray}
Under these conditions, we have the following fluctuation-
dissipation
relation:
\begin{eqnarray}
 \langle {\bf f}({\bf x},t)\rangle &=&0,\nonumber \\
\langle {f}_{i}({\bf x},t){f}_{j}({\bf y},s)\rangle
&=&{k}_{B}T {\eta
}_{ij}({\bf x},{\bf y};t-
s).
\end{eqnarray}
        With a typical Markov approximation, the width in time
of
$\eta_{ij}$ is taken to
be
small,
effectively a delta function.  The fluctuating force is
effectively
white noise, and the
effective Langevin equation can be written:
\FL
\begin{equation}
{\ddot{ x}}_{i}(t)=-{\partial U({\bf x}(t)) \over \partial
{x}_{i}}+{f}_{i}(
{\bf x}(t),t)-{\overline{\eta }}_{ij}({\bf
x}(t)){\dot{x}}_{j}
\end{equation}
and the new fluctuation-dissipation relation is
\begin{equation}
 \langle {f}_{i}({\bf x},t){f}_{j}({\bf x},s)\rangle
={k}_{B}T
{\overline{\eta }}_{ij}({\bf
x})\delta (t-s).\end{equation}

The dependence of the dissipation on spatial correlations of
the
fluctuating forces has been
lost.   Since the particle can only be at one position at one
time,
these correlations
play no role in the classical dynamics.  Indeed, a wide
variety of models can reduce to the same Langevin equation,
if only
in
approximation.
One could even assume a spatially homogeneous environment in
which the
${\overline{\eta }}_{ij}$ is independent
of ${\bf x}$, and the resulting Langevin equation to get
linear
dissipation.
        A quantum mechanical description of the particle and its
dynamics
includes
superpostions between different positions, resulting in
interference
between different
particle
trajectories.  One effect of the interaction with the
oscillator bath is
that the noise will result
in a loss of quantum coherence.  This loss of coherence will
depend
upon
the spatial
correlations of the noise~\cite{Diosi,Gallis}.  In the the
following
sections we will
 explore master equations which have linear dissipation and
include
spatial correlation
effects,
and thus are candidate models for the effects on a quantum
system
resulting from
local coupling to a homogeneous and isotropic environment.

\section{Construction of the master equation}
\label{sec:construct}

        To construct the master equation, we focus on the non
unitary
part
of the evolution,
that is,
\begin{equation}
{ \Delta L}^{{}^*}[O]\equiv {1 \over
2\hbar}\sum\nolimits\limits_{\mu
}^{}
{V}_{\mu }^{\dagger }[O,{V}_{\mu }^{}]+[{V}_{\mu }^{\dagger
},O]{V}_{\mu
}^{}.
\label{dellind}
\end{equation}
The time dependence of expectation values can immediately be
obtained
via
\FL
\begin{equation}
{d\langle O \rangle \over dt}=\langle {\partial O \over
\partial
t}\rangle +\langle {L}^{{}^*}[O]
\rangle = \langle {\partial O
\over \partial t}\rangle +\langle {i \over \hbar}[H,O]\rangle
+\langle
\Delta {L}^{{}^*}[O]\rangle ,
\end{equation}
where $\langle O \rangle ={\rm Tr}[\rho O]$.  We consider
dissipation in
the general case of d
dimensions.
For strictly linear dissipation, we require
\begin{eqnarray}
{ \Delta L}^{{}^*}[{\bf p} ]&=&-{\eta  \over m}{\bf
p},\nonumber \\
{ \Delta L}^{{}^*}[{\bf x}]&=&0,
\label{delclass}
\end{eqnarray}
which will produce quantum Langevin equations corresponding
to
Eq.\
(\ref{class}).
One could also consider more general forms for
the linear dissipation where $\eta$ is replaced with a
dissipation
tensor,
and the classical frictional
force components would be given by $F_i = \eta_{ij} p_j$
(summation
over
$j$), corresponding to an
environment which
was anisotropic.  The anisotropic case could be addressed by
a simple
generalization of
what we present here.

        The operators $\{V_\mu\}$ presumably could be expressed
as
functions of the
position
and momentum operators.  This is the approach taken by
Sandulescu
and
Scutaru~cite{Scutaru}, where
$\{V_\mu\}$ were taken to be linear combinations of position
and
momentum.  We will
choose
the particular form of $\{V_\mu\}$ to be at most linear in
momentum, and
write
\begin{equation}
{V}_{\mu }={A}_{\mu }({\bf x})-{\bf B}_{\mu }({\bf x})\cdot
{\bf p},
\label{vchoice1}
\end{equation}
were $\{{A}_{\mu}({\bf x})\}$ and $\{{\bf B}_{\mu }({\bf
x})\}$ are
as
yet
undetermined functions
of ${\bf x}$.  The choice of this form is
practically dictated by the requirement that the dissipation
be linear,
as any higher powers
of momentum would appear as nonlinear orders of momentum in
the
first
part of
Eq.\ (\ref{delclass}).
We can now use Eq.\ (\ref{dellind}) and Eq.\ (\ref{delclass})
to
establish some conditions on the functions
$\{A_\mu({\bf x})\}$ and $\{B_{\mu i}({\bf x})\}$.
Evaluating Eq.\
(\ref{dellind}) for a
particular
component of momentum, we take
\begin{equation}
{ \Delta L}^{{}^*}[{p}_{n}]={1 \over
2\hbar}\sum\nolimits\limits_{\mu
}^{}
{V}_{\mu }^{}[{p}_{n},{V}_{\mu }^{\dagger }]+[{V}_{\mu
}^{},{p}_{n}]{V}_{\mu
}^{\dagger },\end{equation}
substitute for $\{V_\mu\}$ with Eq.\ (\ref{vchoice1}), and
reorder
the operators with all momentum terms to the right:
\widetext
\begin{eqnarray}
{ \Delta L}^{{}^*}[{p}_{n}]=&&{1 \over
2i}(\sum\nolimits\limits_{\mu
}^{} ({A}_{\mu }^{\dagger }{\partial {A}_{\mu }^{} \over
\partial
{x}_{n}^{}}-
{\partial
{A}_{\mu }^{\dagger } \over \partial {x}_{n}^{}}{A}_{\mu
}^{}))-{\hbar
\over
2}\sum\nolimits\limits_{\mu i}^{} [{\partial  \over \partial
{x}_{i}^{}}({B}_{\mu
i}^{\dagger }{\partial {A}_{\mu }^{} \over \partial
{x}_{n}^{}}-
{\partial {B}_{\mu
i}^{\dagger } \over \partial {x}_{n}^{}}{A}_{\mu
}^{})]\nonumber \\
&&-{\hbar \over 2}\sum\nolimits\limits_{\mu ij}^{}
\{[{\partial
\over
\partial
{x}_{i}^{}}({B}_{\mu i}^{\dagger }{\partial {B}_{\mu j}^{}
\over
\partial
{x}_{n}^{}}-
{\partial {B}_{\mu i}^{\dagger } \over \partial
{x}_{n}^{}}{B}_{\mu
j}^{})]{p}_{j}\}\nonumber \\
&&-{1 \over 2i}\sum\nolimits\limits_{\mu i}^{} [({B}_{\mu
i}^{\dagger
}{\partial
{A}_{\mu
}^{} \over \partial {x}_{n}^{}}-{\partial {B}_{\mu
i}^{\dagger } \over
\partial
{x}_{n}^{}}{A}_{\mu }^{}-{B}_{\mu i}{\partial {A}_{\mu
}^{\dagger }
\over \partial
{x}_{n}^{}}+{\partial {B}_{\mu i} \over \partial
{x}_{n}^{}}{A}_{\mu
}^{\dagger
}){p}_{i}]\nonumber \\
&&+{1 \over 2i}\sum\nolimits\limits_{\mu ij}^{} [({B}_{\mu
i}^{\dagger
}{\partial
{B}_{\mu
j}^{} \over \partial {x}_{n}^{}}-{\partial {B}_{\mu
i}^{\dagger } \over
\partial
{x}_{n}^{}}{B}_{\mu j}^{}){p}_{i}{p}_{j}].
\end{eqnarray}
\narrowtext
The resulting expression contains terms up to second order in
momentum,
and we equate
each power to the corresponding term in Eq.\
(\ref{delclass}),
which produces, after some simplification, the following
conditions
on  $\{A_\mu({\bf x})\}$ and $\{B_{\mu i}({\bf x})\}$:
\widetext
\begin{eqnarray}
\sum\nolimits\limits_{\mu ij}^{} {B}_{\mu i}^{^{*}}{\partial
{B}_{\mu
j}^{} \over \partial {x}_{n}^{}}-{\partial {B}_{\mu i}^{^{*}}
\over
\partial
{x}_{n}^{}}{B}_{\mu j}^{}&=&0\nonumber \\
{1 \over 2i}\sum\nolimits\limits_{\mu i}^{} ({B}_{\mu
i}^{^{*}}{\partial
{A}_{\mu
}^{} \over \partial {x}_{n}^{}}-{\partial {B}_{\mu i}^{^{*}}
\over
\partial
{x}_{n}^{}}{A}_{\mu }^{}-{B}_{\mu i}{\partial {A}_{\mu
}^{^{*}} \over
\partial
{x}_{n}^{}}+{\partial {B}_{\mu i} \over \partial
{x}_{n}^{}}{A}_{\mu
}^{^{*}
})&=&{\delta }_{in}{\eta  \over m}\nonumber \\
{1 \over 2i}\sum\nolimits\limits_{\mu }^{} ({A}_{\mu
}^{^{*}}{\partial
{A}_{\mu
}^{} \over \partial {x}_{n}^{}}-{\partial {A}_{\mu }^{^{*}}
\over
\partial
{x}_{n}^{}}{A}_{\mu }^{})+{\hbar \over
2}\sum\nolimits\limits_{\mu
i}^{}
[{\partial
\over \partial {x}_{i}^{}}({B}_{\mu i}^{^{*}}{\partial
{A}_{\mu }^{}
\over \partial
{x}_{n}^{}}-{\partial {B}_{\mu i}^{^{*}} \over \partial
{x}_{n}^{}}{A}_{\mu
}^{})]&=&0.
\label{cond1a}
\end{eqnarray}
\narrowtext
To satisfy the second part of Eq.\ (\ref{delclass}), we look
at Eq.\
(\ref{dellind})
 for a particular component
of the position operator to get
\begin{eqnarray}
{ \Delta L}^{{}^*}[{x}_{n}]&=&{1 \over
2}\{\sum\nolimits\limits_{\mu
j}^{} [-\hbar{\partial  \over \partial {x}_{j}^{}}({B}_{\mu
j}^{\dagger
}{B}_{\mu
n})]\nonumber \\
&&+{1 \over i}\sum\nolimits\limits_{\mu }^{} [({A}_{\mu
}^{\dagger
}{B}_{\mu n}^{}-
{A}_{\mu }^{}{B}_{\mu n}^{\dagger }]\nonumber \\
&&+{1 \over i}\sum\nolimits\limits_{\mu j}^{} [{B}_{\mu
n}^{\dagger
}{B}_{\mu j}-
{B}_{\mu j}^{\dagger }{B}_{\mu n}){p}_{j}]\}.
\end{eqnarray}
The resulting constraint equations are
\FL
\begin{eqnarray}
\sum\nolimits\limits_{\mu j}^{} [-\hbar{\partial  \over
\partial
{x}_{j}^{}}({B}_{\mu j}^{^{*}}{B}_{\mu n})]+{1 \over
i}\sum\nolimits\limits_{\mu
}^{} [({A}_{\mu }^{^{*}}{B}_{\mu n}^{}-{A}_{\mu }^{}{B}_{\mu
n}^{^{*}
}]=0\nonumber \\
\sum\nolimits\limits_{\mu j}^{} [{B}_{\mu n}^{^{*}}{B}_{\mu
j}-
{B}_{\mu
j}^{^{*}}{B}_{\mu n})]=0.
\label{cond1b}
\end{eqnarray}

        The conditions above by no means determine the functions
$\{A_\mu({\bf x})\}$
and
$\{B_{\mu i}({\bf x})\}$, but are constraints to be
satisfied.  The
master
equation we are generating is presumably a limiting form of
the
reduced
dynamics of a
composite quantum system, so we select a particular form of
$\{A_\mu({\bf x})\}$ and
$\{B_{\mu i}({\bf x})\}$ to reflect the form of the
interaction
between
the system of
interest
and the environment degrees of freedom.  For a local
interaction with
a
field, the
environment degrees of freedom are the particular modes of
the
field.  A
reasonable choice
for these functions are then
\begin{eqnarray}\{{A}_{\mu }\}&=&\{a({\bf k}){e}^{i{\bf
k}\cdot {\bf
x}}\}\nonumber \\
\{{B}_{\mu j}\}&=&\{{b}_{j}({\bf k}){e}^{i{\bf k}\cdot {\bf
x}}\},
\end{eqnarray}
where the index $\mu$ is replaced by a continuous $d$
dimensional
wave
vector ${\bf
k}$,
and sums
over $\mu$ are replaced by integrations over ${\bf k}$.  Some
simplification occurs
immediately
since, with this choice, terms such as the product
$A_m^{^{*}} B_{mi}$
are independent
of ${\bf x}$.
The conditions expressed in Eq.\ (\ref{cond1a}) and Eq.\
(\ref{cond1b})
reduce to
\begin{eqnarray}\int_{ }^{}{d}^{d}k{b}_{i}^{^{*}}({\bf
k}){b}_{j}({\bf
k}){k}_{n}&=&0\nonumber \\
\int_{}^{}{d}^{d}k2{\rm Re}({a}^{^{*}}({\bf k}){b}_{i}({\bf
k})){k}_{n}&=&{\delta
}_{in}{\eta  \over
m}\nonumber \\
\int_{}^{}{d}^{d}k|{a}^{^{*}}({\bf
k}){|}^{2}{k}_{n}&=&0\nonumber \\
\int_{}^{}{d}^{d}k{\rm Im}({a}^{^{*}}({\bf k}){b}_{i}({\bf
k}))&=&0\nonumber \\
\int_{}^{}{d}^{d}k{\rm Im}({b}_{j}^{^{*}}({\bf
k}){b}_{i}({\bf k}))&=&0.
\end{eqnarray}
To satisfy these conditions, we need $a({\bf k})$ to be an
even
function
of the components of ${\bf
k}$,
and $b_i({\bf k})$ to be an odd function of $k_i$, but an
even
function
of the other
components of ${\bf k}$.
To insure isotropy, we choose the form of the functions
$a({\bf k})$
and
$a({\bf k})$
 as
\begin{eqnarray}a({\bf k})&=&\alpha (k)\nonumber \\
{b}_{j}({\bf k})&=&\beta (k){k}_{j}.
\end{eqnarray}
These functions can then be seen to satisfy the conditions
(Eqs.
(\ref{cond1a})
and (\ref{cond1a})) by considering the symmetry of
the even limits for the integrals over particular components
of ${\bf
k}$, along with
 the even or odd integrands.  The dampening constant is given
for
this
final choice by
\begin{eqnarray}
 {\eta  \over m}&=&\int_{}^{}{d}^{d}k2{\rm Re}({\alpha
}^{{}^*}(k)\beta
(k)){k}_{n}^{2} \nonumber \\
&=&\int_{}^{}{d}^{d}k2{\rm Re}({\alpha }^{{}^*}(k)\beta
(k)){{k}^{2} \over d}.
\end{eqnarray}

        We can now write the nonunitary part of the evolution
equations
for the particular
choice of $\{V_\mu \}$. For the Heisenberg picture (after
some
simplification using the
symmetries
of the functions $\alpha$ and $\beta$) we have for an
arbitrary
operator
$O$,
\widetext
\begin{eqnarray}
{\Delta L}^{{}^*}[O]=&&\int_{}^{}{d}^{d}k{|\alpha (k){|}^{2}
\over
\hbar}({e}^{-i{\bf k}\cdot {\bf x}} O{e}^{i{\bf k}\cdot {\bf
x}}-
O)\nonumber \\
&+&\int_{}^{}{d}^{d}k{|\beta (k){|}^{2} \over \hbar}({\bf
k}\cdot {\bf
p} e^{-i{\bf
k} \cdot {\bf x}} O{e}^{i{\bf k}\cdot {\bf x}}k \cdot {\bf p}
-{1 \over
2}\{O,({\bf
k} \cdot {\bf p} )^{2}\})\nonumber \\
&-&\int_{}^{}{d}^{d}k{{\rm Re}(\alpha (k{)}^{{}^*}\beta (k))
\over
\hbar}(\{{e}^{-
i{\bf k}
\cdot {\bf x}} O{e}^{i{\bf k}\cdot {\bf x}} ,{\bf k}\cdot
{\bf p}
\})\nonumber \\
&-&\int_{}^{}{d}^{d}k{i{\rm Im}(\alpha (k{)}^{{}^*}\beta (k))
\over
\hbar}([{e}^{-
i{\bf k}
\cdot {\bf x}} O{e}^{i{\bf k}\cdot {\bf x}},{\bf k}\cdot {\bf
p} ]),
\label{delh}
\end{eqnarray}
while for the Schr\"odinger picture we have
\begin{eqnarray}\Delta L[\rho ]&=&-\int_{}^{}{d}^{d}k{|\alpha
(k){|}^{2}
\over
\hbar}(\rho
-{e}^{i{\bf k}\cdot {\bf x}} \rho {e}^{-i{\bf k}\cdot {\bf
x}}
)\nonumber \\
&&-\int_{}^{}{d}^{d}k{|\beta (k){|}^{2} \over \hbar}({1 \over
2}\{({\bf
k}\cdot {\bf
p}{
)}^{2},\rho \}-{\bf k}\cdot {\bf p} e^{i{\bf k}\cdot {\bf x}}
\rho e^{-
i{\bf k}
\cdot {\bf x}}{\bf k} \cdot {\bf p} )\nonumber \\
&&-\int_{}^{}{d}^{d}k{{\rm Re}(\alpha (k{)}^{{}^*}\beta (k))
\over
\hbar}(\{{\bf
k}\cdot {\bf
p} ,e^{i{\bf k}\cdot {\bf x}} \rho e^{-i{\bf k}\cdot {\bf x}}
\})\nonumber \\
&&-\int_{}^{}{d}^{d}k{i{\rm Im}(\alpha (k{)}^{{}^*}\beta (k))
\over
\hbar}([{\bf
k}\cdot {\bf
p},e^{i{\bf k}\cdot {\bf x}} \rho {e}^{-i{\bf k}\cdot {\bf
x}} ]).
\label{dels}
\end{eqnarray}
\narrowtext
This is the main result of this paper: we have illustrated a
method of
constructing
quantum master equations which are ohmic and
translationally invariant.  The translational invariance can
easily be
seen by
replacing ${\bf x}$ with
${\bf x}'+{\bf r}$, where ${\bf r}$ is a c-number (${\bf p}$
is
unchanged by
translation).  The form of the nonunitary
terms is unchanged.

\section{Dissipation, fluctuation \\ and expectations}
\label{sec:dissipation}

        In this section, we will examine the implications of the
dissipative part of the
evolution given by Eq.\ (\ref{delh}) or Eq.\ (\ref{dels}). by
making
use
of the
Heisenberg evolution equations,
where the time dependence of expectation values is given by:
\begin{equation}
 {d\langle O \rangle \over dt}={\rm Tr}[\rho
({L}^{{}^*}[O]+{\partial O
\over \partial t})].
\label{expo}
\end{equation}
 Much immediate simplification can occur when one notes that
$e^{i{\bf
k}\cdot {\bf x}}
 O e^{-i{\bf k}\cdot {\bf x}}$ is simply a
unitary transformation, corresponding to a translation of
$\hbar {\bf
k}$  in momentum
space.
Additional simplifications occur with the symmetries of the
integrals
over k and the
particular form of $\alpha$ and $\beta$ as function of
$k=|{\bf k}|$.
{}From this, one can
readily
 verify that $\Delta L^{{}^*}({\bf p}) = - {\eta \over m}{\bf
p}$.

We now wish to look at the effect of the non unitary part of
the
evolution on kinetic
energy.  We find, with some simplification which arises from
the
isotropy of the model,
that
\begin{eqnarray}
 \Delta {L}^{{}^*}[{{P}^{2} \over 2m}]
&=&\int_{}^{}{d}^{d}k{|\alpha
(k){|}^{2}
\over 2m}\hbar{k}^{2} \nonumber \\
&&+(\int_{}^{}{d}^{d}k{|\beta (k){|}^{2} \over
d}\hbar{k}^{4}-
2{\eta  \over m}){{P}^{2} \over 2m}.
\label{delke1}
\end{eqnarray}
This can be interpreted as a relaxation to thermal
equilibrium, under
certain circumstances.
We can make the identifications
\begin{eqnarray}\gamma &\equiv& 2{\eta  \over m}-
\int_{}^{}{d}^{d}k{|\beta
(k){|}^{2}
\over d}\hbar{k}^{4}\nonumber \\
{d \over 2}{k}_{B}T &\equiv& {1 \over \gamma
}\int_{}^{}{d}^{d}k{|\alpha
(k){|}^{2}
\over 2m}\hbar{k}^{2},
\label{param}
\end{eqnarray}
which determine an effective equilibrium temperature,
provided
$\gamma >
0$.  Since
$\Delta L^{{}^*}$ is linear
and since $\Delta L^{{}^*}({\rm c-number}) = 0$, we can now
rewrite
Eq.\
(\ref{delke1}) as
\begin{equation}
{ \Delta L}^{{}^*}[({{P}^{2} \over 2m}-{d \over
2}{k}_{B}T)]=-\gamma
({{P}^{2} \over 2m}-{d \over 2}{k}_{B}T).
\label{delke2}
\end{equation}

        The first term on the RHS of Eq.\ (\ref{dels}) can be
identified
as the effect of a
random
gaussian fluctuating potential.  For a particle in a random
potential
with a translationally
invariant two point correlation function with short
relaxation time,
approximated by
\begin{equation}
\langle V({\bf x},t)V({\bf y},s)\rangle =g({\bf x}-{\bf
y})\delta (t-
s),\end{equation}
the master equation gains an additional term which
is given in the position representation
by~\cite{Diosi,Gallis}
\begin{equation}
 \langle {\bf x}|\Delta L(\rho )|{\bf y}\rangle =-{1 \over
{\hbar}^{2}}(g(0)-g({\bf x}-\bf
y ))\rho ({\bf x},{\bf y}).\end{equation}
The position representation of the first term on the RHS of
Eq.\
(\ref{dels}) is then given
by
\FL
\begin{eqnarray}
&-& \langle {\bf x}|\int_{}^{}{d}^{d}k{|\alpha (k){|}^{2}
\over
\hbar}(\rho -{e}^{i\bf
k \cdot \bf x} \rho {e}^{-i{\bf k}\cdot \bf x} )|{\bf
y}\rangle
\nonumber \\
=&-& \int_{}^{}{d}^{d}k{|\alpha (k){|}^{2} \over \hbar}(1-
{e}^{i{\bf
k}\cdot ({\bf x}-
\bf
y )})\rho (x,y) ,
\end{eqnarray}
so that after some simplification, we can identify this as
the effect of
a random gaussian
potential with correlation function given by
\FL
\begin{eqnarray}
 \langle V({\bf x},t)V({\bf y},s)\rangle =\nonumber \\
\hbar\int_{}^{}{d}^{d}k|\alpha ({\bf k}
){|}^{2}\cos({\bf k}\cdot (x-y))\delta (t-s).
\end{eqnarray}
In terms of fluctuating forces, we can write
\begin{eqnarray}
\langle {\bf f}({\bf x},t)\cdot {\bf f}({\bf y},s)\rangle =
{\nabla }_{x}\cdot {\nabla }_{y}\langle V({\bf x}
,t)V({\bf y},s)\rangle \nonumber \\
= \hbar\int_{}^{}{d}^{d}k|\alpha (k){|}^{2}{k}^{2}\cos({\bf
k}\cdot
({\bf x}-{\bf y}))\delta (t-s).
\end{eqnarray}
The fluctuation-dissipation relation can now be written using
the
identifications made in
Eq.\ (\ref{param}),
\FL
\begin{eqnarray}
 \langle {\bf f}({\bf x},t)\cdot {\bf f}({\bf x},s)\rangle
&=&
\langle {\bf f}(0,t)\cdot {\bf f}(0,s)\rangle \nonumber \\
&=& 2m\gamma {d \over 2}{k}_{B}T\delta (t-s).
\label{flucdis}
\end{eqnarray}

We now wish to consider the expectations of more general
quantities.
For an
 arbitrary function of the position operator $g({\bf x})$, it
is
straightforward to examine the
time dependence of its expectation value via Eq.\
(\ref{expo}).  Since
x
is unaffected by
translations in momentum space,  evaluating $\Delta
L^{{}^*}[g({\bf
x})]$
via Eq.\ (\ref{delh}) becomes
\FL
\begin{eqnarray}
{ \Delta L}^{{}^*}[g({\bf x})]&=&\nonumber \\
&&+\int\nolimits {d}^{d}k{{|\beta (k)|}^{2} \over \hbar}({\bf
k}\cdot
{\bf p} g({\bf
x}){\bf
k} \cdot {\bf p} -{1 \over 2}\{g({\bf x}),({\bf k}\cdot {\bf
p}{)}^{2}\})\nonumber \\
&&-\int\nolimits {d}^{d}k{{\rm Re}(\alpha (k{)}^{*}\beta (k))
\over
\hbar}(\{g({\bf
x}),\bf
k \cdot {\bf p} \})\nonumber \\
&&-\int\nolimits{d}^{d}k{i{\rm Im}(\alpha (k{)}^{*}\beta (k))
\over
\hbar}([g({\bf
x}),{\bf k} \cdot {\bf p} ]),
\end{eqnarray}
which is further simplified via the symmetries of the
integrals over
${\bf k}$ to
\FL
\begin{eqnarray}
{ \Delta L}^{{}^*}[g({\bf
x})]&=&\int_{}^{}{d}^{d}k{\hbar{k}^{2}|\beta
(k){|}^{2} \over 2d}{\nabla }^{2}g({\bf x}) \nonumber \\
&\equiv& {D \over 2}{\nabla }^{2}g({\bf x}
),\end{eqnarray}
effectively defining the diffusion constant $D$.  The
spreading of the
density operator can
then  be characterized by
\begin{equation}
{ d\langle {({\bf x}-{\bf x}_{ o})}^{2}\rangle \over
dt}=\langle {i
\over \hbar}[H,{({\bf x}
-{\bf x}_{ o})}^{2}]\rangle +D.\end{equation}
For an arbitrary function of momentum, $h({\bf p})$ we have
\begin{equation}
e^{i{\bf k} \cdot {\bf x}}h({\bf p}) e^{-i{\bf k} \cdot {\bf
x}}  =
h({\bf p}+\hbar {\bf
k}),\end{equation}
and $\Delta L^{{}^*}[h({\bf p})]$  becomes
\begin{eqnarray}
{ \Delta L}^{{}^*}[h({\bf p} )]= \nonumber \\
\int_{}^{}{d}^{d}k{|\alpha (k){|}^{2}
\over \hbar}(h({\bf p} +\hbar{\bf k})-h({\bf p} ))\nonumber
\\
+\int_{}^{}{d}^{d}k{|\beta (k){|}^{2} \over \hbar}[({\bf
k}\cdot {\bf
p}{
)}^{2}(h({\bf p} +\hbar{\bf k})-h({\bf p} ))]\nonumber \\
-\int_{}^{}{d}^{d}k{2{\rm Re}(\alpha (k{)}^{{}^*}\beta (k))
\over
\hbar}({\bf
k}\cdot {\bf p})h({\bf p} +\hbar{\bf k}).
\end{eqnarray}

\section{Comparison with Dekker and \\ Caldeira-Leggett
Master
Equations}
\label{sec:compare}

        It is useful to take a look at Sandulescu and Scutaru's
construction of the Dekker
master equation  in order to make comparisons with our
results.  In
section two, we were
largely followed their construction.  Their choice for the
operators
$\{V_\mu \}$ were
linear
combinations of position and momentum operators,
\begin{equation}
{ V}_{i}={a}_{i}x+{b}_{i}p,\end{equation}
while in our d-dimensional generalization, we considered
operators
of
the form:
\begin{equation}
{ V}_{\bf k} =\alpha ({\bf k}){e}^{i{\bf k}\cdot \bf x}
-\beta ({\bf k}
){e}^{i{\bf k}\cdot \bf x}{\bf k} \cdot {\bf p} .
\label{vchoicef}
\end{equation}
It is possible to consider the Dekker equation as a low
length scale
limit of our result.
One can imagine that this will correspond to a system which
is well
localized, perhaps by
the
potential of the system of interest.  We can take the origin
to be
centered in the region of
localization, noting that translations will not affect the
form of
$\Delta L$.
        The length scale is determined the physical extent of
this
region
of
localization.  A short length scale approximation can be
considered as
a
long wavelength
(for
the environment modes) approximation via the "typical" wave
numbers in
Eq.
(\ref{vchoicef}).
To make this approximation we expand in powers of $k$.  To
second
order,
we have
\FL
\begin{equation}
{ e}^{i{\bf k}\cdot \bf x} O{e}^{-i{\bf k}\cdot \bf x} \cong
O+i[{\bf k}
\cdot {\bf x},O]-{1 \over 2}[{\bf k}\cdot {\bf x},[{\bf
k}\cdot {\bf
x},O]].\end{equation}
With simplifications (due to the symmetry in our choice of
$\alpha$
and
$\beta$),
 $\Delta L$ becomes
\FL
\begin{eqnarray}
\Delta L[\rho ]&\cong& -\int_{}^{}{d}^{d}k{|\alpha (k){|}^{2}
\over
2\hbar}[{\bf k}\cdot {\bf x},[{\bf k}\cdot {\bf x},\rho
]]\nonumber \\
&&-\int_{}^{}{d}^{d}k{|\beta (k){|}^{2} \over 2\hbar}[{\bf
k}\cdot {\bf
p} ,[{\bf k}
\cdot {\bf p} ,\rho ]]\nonumber \\
&&-\int_{}^{}{d}^{d}k{i{\rm Re}(\alpha (k{)}^{{}^*}\beta (k))
\over
\hbar}[{\bf
k}\cdot \bf
x ,\{{\bf k}\cdot {\bf p} ,\rho \}]\nonumber \\
&&+\int_{}^{}{d}^{d}k{{\rm Im}(\alpha (k{)}^{{}^*}\beta (k))
\over
\hbar}[{\bf
k}\cdot \bf
x ,[{\bf k}\cdot {\bf p} ,\rho ]].
\end{eqnarray}
In 1 dimension, this corresponds to the Dekker master
equation, with
\begin{eqnarray}
 \Delta L[\rho ]&=&-i{\lambda  \over \hbar}[x,\{p,\rho \}]+{1
\over
{\hbar}^{2}}({D}_{xp}+{D}_{px})[x,[p,\rho ]]\nonumber \\
&&-{{D}_{xx} \over
{\hbar}^{2}}[p,[p,\rho ]]-{{D}_{pp} \over
{\hbar}^{2}}[x,[x,\rho ]],
\end{eqnarray}
where the parameters are given by
\begin{eqnarray}{ D}_{pp}&=&\int_{}^{}{d}^{d}k {\hbar|\alpha
(k){|}^{2}
\over
2}{k}^{2}\nonumber \\
{D}_{xx}&=&\int_{}^{}dk {\hbar|\beta (k){|}^{2} \over
2}{k}^{2}\nonumber
\\
\lambda &=&\int_{}^{}dk \, i {\rm Re}(\alpha
(k{)}^{{}^*}\beta
(k)){k}^{2}={\eta  \over
2m}\nonumber
\\
({D}_{xp}+{D}_{px})&=&\int_{}^{}dk\, \hbar {\rm Im}(\alpha
(k{)}^{{}^*}\beta
(k)){k}^{2}.
\end{eqnarray}

        For the Caldeira-Leggett master equation, we have
\begin{equation}
 \Delta L[\rho ]={\eta  \over i2m\hbar}[x,\{p,\rho \}]-{\eta
{k}_{B}T
\over
{\hbar}^{2}}[x,[x,\rho ]],\end{equation}
which can be viewed as a low momentum  approximation
(ignoring
second
order
momentum
terms
in $\Delta L$) of the Dekker master equation with $D_{xp} =
D_{px} =
0$.
If we make similar
approximations for our master equation, without making short
length
scale
approximations, we have
\FL
\begin{eqnarray}
\Delta L[\rho ]=-\int_{}^{}{d}^{d}k{|\alpha (k){|}^{2} \over
\hbar}(\rho
-{e}^{i{\bf k}\cdot {\bf x}} \rho {e}^{-i{\bf k}\cdot {\bf
x}}
)\nonumber \\
-\int_{}^{}{d}^{d}k{{\rm Re}(\alpha (k{)}^{{}^*}\beta (k))
\over
\hbar}(\{{\bf
k}\cdot {\bf
p},{e}^{i{\bf k}\cdot {\bf x}} \rho {e}^{-i{\bf k}\cdot {\bf
x}}\}).
\end{eqnarray}
For this case, the fluctuation-dissipation relations are of a
more
familiar form, relating the
effective temperature to the correlation of the fluctuating
potentials,
where the equivalent
relations to Eqs.\ (\ref{param}) and(\ref{flucdis}) become
\begin{eqnarray}
\gamma &=&2{\eta  \over m}\nonumber \\
{d \over 2}{k}_{B}T&\equiv& {1 \over \gamma
}\int_{}^{}{d}^{d}k{|\alpha
(k){|}^{2}
\over 2m}\hbar{k}^{2},
\end{eqnarray}
and
\begin{eqnarray}\langle {\bf f}(x,t) \cdot {\bf
f}(x,s)\rangle &=&
\langle {\bf f}(0,t)\cdot {\bf f}(0,s)\rangle \nonumber \\
&=& 2m\gamma {d \over 2}{k}_{B}T\delta (t-s)\nonumber \\
&=& 4 \eta {d \over 2}{k}_{B}T\delta (t-s).\end{eqnarray}

\section{Discussion}
\label{sec:discuss}

        We have introduced a method of construction of models
for
linear
(i.e. ohmic)
dissipation for a quantum system for arbitrary dimension.
The
model is
completely
specified by identifying the complex functions $a({\bf k})$
and ${\bf
b}({\bf k})$.
 Although we have focused on isotropic
dissipation, anisotropic dissipation can readily be
introduced by
retaining the direction
dependence of $a$ and $b$ on ${\bf k}$.  We have identified
the
phenomenological dissipation constant, and effective
temperature, as
well as the spatial
correlation function for the fluctuating forces.
Generalizations of
this approach might
also include considering other sets of non-trivial functions
for
$\{A_\mu({\bf x})\}$ and $\{B_{\mu i}({\bf x})\}$, so that
the
conditions expressed
in Eqs.\ (\ref{cond1a}) and (\ref{cond1b}) are satisfied when
averaged
over some
characteristic length scale.

        The models we have introduced are natural
generalizations of
the
Dekker master
equation, and in one dimension, reduce to it in a short
length scale
approximation.
Conceivable, one could construct a more general master
equation
using
Dekker's noise
operators by including spatial correlation effects.  The
Caldeira-
Leggett master equation can
be viewed as a short-length scale, small momentum
approximation to
our
result.

         Caldeira and Leggett have pointed out that the only
coupling to
an oscillator bath
which {\it exactly} produces ohmic dissipation is linear
coupling
(in both system and environment coordinates).   However, one
might
expect that nonlinear coupling may provide dissipation which
is
linear
under a wide range of
approximations.  The models we have proposed here are
intended to
anticipate the
form of the dynamics which would result from local coupling
to an
environment.
We have previously argued that the
effect on the dynamics of quantum coherence may differ
considerably from
that which can be
obtained by linear coupling.~\cite{GF,Gallis}

\acknowledgements

I would like to acknowldge the hospitality of the Aspen
Center for
Physics,
where some of where part of this work was done.  I am also
grateful
to
Salman
 Habib for several enlightening conversations.

\end{document}